# Unlocking device-scale atomistic modelling of phase-change memory materials

Yuxing Zhou,[1,2] Wei Zhang,[2,*] En Ma,[2] and Volker L. Deringer[1,*]

[1]*Department of Chemistry, Inorganic Chemistry Laboratory, University of Oxford, Oxford OX1 3QR, United Kingdom*

[2]*Center for Alloy Innovation and Design (CAID), State Key Laboratory for Mechanical Behavior of Materials, Xi'an Jiaotong University, Xi'an 710049, China*

*E-mail: wzhang0@mail.xjtu.edu.cn; volker.deringer@chem.ox.ac.uk

**Quantum-accurate computer simulations play a central role in understanding phase-change materials (PCMs) for advanced memory technologies. However, direct quantum-mechanical simulations are necessarily limited to simplified models, containing no more than a few hundred or a thousand atoms. Machine learning (ML) based potential models that are "trained" on quantum-mechanical data are an emerging alternative approach, currently evolving from highly specialised to more widely applied simulation tools. Here we show that a universal, compositionally flexible ML model can describe a wide range of flagship Ge–Sb–Te PCMs under real device conditions, including non-isothermal heating and chemical disorder which are relevant for memory applications. The speed of the ML model enables atomistic simulations of multiple thermal cycles and delicate operations for neuro-inspired computing, namely, cumulative SET and iterative RESET. A device-scale capability demonstration (40 × 20 × 20 nm³) shows that the new ML potential can directly describe technologically relevant processes in PCM-based memory products. In a wider context, our work demonstrates how ML-driven materials simulations are now entering a stage where they can guide architecture design for high-performance electronic devices.**

**Introduction**

Phase-change material (PCM) based random access memory is a leading candidate for non-volatile memory and neuro-inspired computing technologies.[1-4] PCMs can be switched reversibly and rapidly between two solid-state phases, via crystallisation (SET) and amorphisation (RESET), respectively. A large property contrast between the amorphous and crystalline phases, both in electrical conductivity and optical reflectivity, is used to encode digital information. The most useful PCMs are located on the quasi-binary GeTe–$Sb_2Te_3$ line of chemical compositions ("GST"). For example, $Ge_2Sb_2Te_5$ and $Ge_8Sb_2Te_{11}$ are used in re-writeable optical disks,[1] $Ge_1Sb_2Te_4$ has been claimed as the key ingredient for the 3D Xpoint™ memory products,[5] and new $Sb_2Te_3$-based PCMs with suitable dopants or nanoconfinement layers are promising candidates for universal memory[6-8] or neuro-inspired computing devices.[9] When integrated with waveguides, $Ge_2Sb_2Te_5$ enables various on-chip non-volatile photonic applications, including memory, computing, colour rendering, and nanopixel displays.[10]

Computer simulations based on density functional theory (DFT) and DFT-driven *ab initio* molecular dynamics (AIMD) have provided key insights into structural features,[11-13] bonding contrast,[14,15] and crystallisation kinetics of GST.[16-19] However, the high computational cost of AIMD prevents simulations of systems beyond a thousand atoms or so, and of processes taking more than a few nanoseconds;[19] at the same time, the structural and chemical complexity in PCMs calls for quantum-mechanically accurate methods rather than empirically parameterised force-field models. These requirements have made it challenging, and often impossible, to simulate more realistic scenarios encountered in devices: Joule heating induces a temperate gradient by programming pulses; the active volume expands and shrinks during switching; the local chemical composition evolves; and model system sizes of tens of nanometres are required to account for the "real-world" device geometry.

Machine learning (ML) based interatomic potentials are an emerging approach to address the challenges mentioned above, aiming to combine the efficiency of empirical potentials and the accuracy of DFT.[20-22] Based on a reference database of small-scale DFT computations, the potential energy surface (PES) becomes the regression target for an ML model: hence, the information about atomic interactions is extracted from high-quality reference data, rather than encoded in the model from the start. In 2012, Sosso *et al.* developed an artificial-neural-network ML potential for the prototype PCM, GeTe,[23] which has since enabled studies of crystallisation,[24] ageing,[25] and thermal transport.[26] More recently, neural-network potentials for elemental Sb ("monatomic memory") were reported.[27,28] In 2018, an ML potential for the ternary flagship PCM, $Ge_2Sb_2Te_5$, was fitted using the Gaussian approximation potential (GAP) framework,[29] and subsequently applied to assess structural properties and the mid-gap electronic states of the amorphous "zero bit".[30,31] However, thus far, these models had been designed and verified for specific compositions, and whilst the extrapolation of the aforementioned GAP to binary $Sb_2Te_3$ has been explored,[32] a comprehensive model for the entire GeTe–$Sb_2Te_3$ quasi-binary line has remained an outstanding challenge. One of the reasons will surely be the sheer structural and chemical complexity that needs to be "taught" to an ML potential: for example, state-of-the-art general-purpose GAPs even for single elements, *viz.* silicon and phosphorus,[33,34] are based on reference databases of more than 100,000 individual atomic environments.

In the present work, we describe a qualitative step forward in the atomistic modelling of PCMs, demonstrating that fully atomistic device-scale simulations are made possible by ML approaches. We show that ML-driven modelling can address questions that are key in novel PCM architectures, such as the simulation of cumulative SET and iterative RESET processes—thereby revealing the atomistic processes and mechanisms in PCMs on the full ten-nanometre length scale that is relevant to devices. More generally, this work illustrates the emerging role of ML-driven simulations in materials science and device engineering.

## Results

**ML-driven modelling of Ge–Sb–Te PCMs.** The quality of any ML model hinges on the quality of its input data, and the performance of an ML potential depends in no small parts on the quality of the reference database to which it is fitted.[35] We here introduce a dataset of representative structural models (data locations) and DFT-computed energies and forces (data labels) that enables the fitting of an ML potential for GST PCMs. Starting with an initial set of crystalline models with defects and structural distortions, and various liquid and amorphous configurations from AIMD (iteration zero, "iter-0"), we iteratively explored structural space with evolving ML potentials and progressively trained the models to represent more complex scenarios (Fig. 1a). Iterative training allows for efficient reference data generation,[35] because no further AIMD is required beyond the initial set of data, which we call "iter-0". We started by performing GAP-driven melt–quench simulations, of the type that are common in the field ("iter-1"), and then followed with a series of more specialised simulations, which we designed to reflect the particular requirements of the domain of application ("iter-2"). The latter allow the model to "learn" microscopic information that is most relevant to GST, *i.e.*, about melting and crystallisation processes in which phases with different degrees of structural order coexist. The "iter-2" part is central here, because it gradually explores and includes the domain knowledge that is relevant to how the PCM functions and evolves in a real-world action.

We illustrate the composition of the reference database in Fig. 1b. In this two-dimensional structural map, points represent individual simulation cells, and the distance between any two points indicates the SOAP kernel similarity[36] of the corresponding configurations. The areas labelled (1)–(3) represent crystalline structures included from the start (*grey*): the first two comprise ordered and disordered GST, respectively, for which the key structural motif is an (often defective) octahedral environment, and therefore these points are distinct from area (3), representing crystalline Ge with its tetrahedral motif. Area (4) covers liquid and amorphous GST configurations, mostly added during early iterations (*yellow*); area (5) reflects phase-transition snapshots, added to the database during the domain-specific series (*red*), and accordingly the points are located between the fully ordered and fully disordered regions of the map.

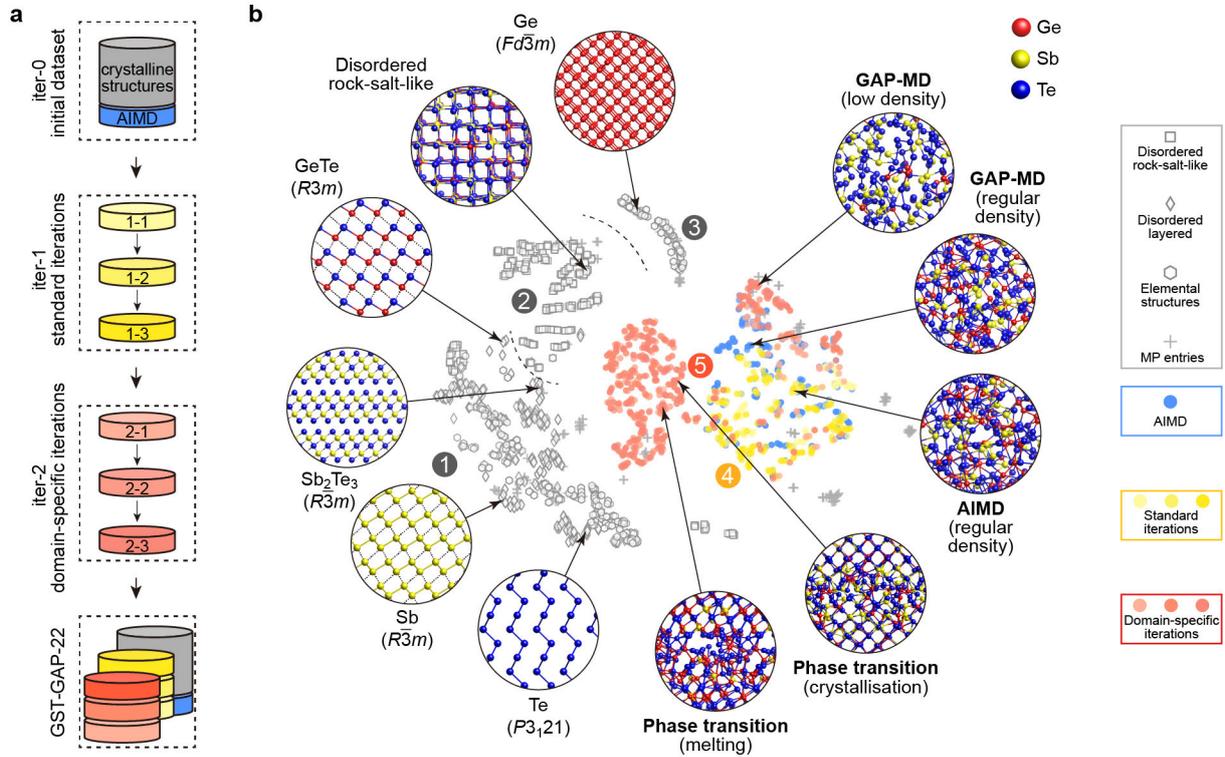

**Figure 1. A machine-learned potential for Ge–Sb–Te phase-change materials.** (**a**) Reference data generation for increasingly comprehensive GAP ML models. The initial reference database ("iter-0") contains crystalline structures and *ab initio* MD (AIMD) snapshots of liquid and amorphous phases. A two-step iterative training process is then carried out, as described in the text, with standard iterations ("iter-1") and domain-specific iterations ("iter-2") gradually expanding the database. (**b**) A two-dimensional structural map representing the reference database, generated using the SOAP similarity matrix (Methods). Representative structures are shown, and are discussed in the text according to the labels (1)–(5).

With the full database available, we fitted a final ML model, to which we refer as GST-GAP-22 in the following. Details of the model hyperparameter and reference-point selection are given in Supplementary Note 1. We assessed the numerical performance of this potential both by cross-validation on the database itself, and by testing on out-of-sample configurations generated separately *via* AIMD. The error analysis suggests that the model has indeed "learned" its quantum-mechanical reference data largely to within the regularisation (the imposed uncertainty during the fit; Supplementary Information). We fitted GST-GAP-22 to reference data obtained with the PBEsol exchange–correlation functional, but we also show that a version based on a different functional can be generated by re-computing the DFT data that "label" the database entries; this is described in the Supplementary Information.

The resulting ML model accurately predicts the structural properties of GST materials. For crystalline phases, GAP-optimised lattice parameters are nearly identical with DFT data (Supplementary Table 2). For disordered structures, Fig. 2a shows that the radial distribution function (RDF) and angular distribution function (ADF) profiles of various liquid and amorphous GST phases are faithfully described. Note that the composition $Ge_4Sb_2Te_7$ (formally "4 GeTe + 1 $Sb_2Te_3$") was not explicitly included during training, yet key structural features are well reproduced by the ML potential. Figure 2b illustrates typical local bonding patterns in amorphous GST: some Ge atoms form tetrahedral motifs, and the majority of Ge atoms (and almost all Sb atoms) form defective octahedral ones.[11,12] All amorphous models contain a non-negligible fraction of homopolar bonds (absent in the respective crystals), *viz.* Ge–Ge, Sb–Sb, Ge–Sb, and Te–Te (Fig. 2c), typically formed during rapid quenching. The fraction of tetrahedral motifs is higher when PBE is used rather than PBEsol, both for AIMD and as a basis for GAP-MD, as a direct consequence of the different types of reference data (Supplementary Fig. 4).

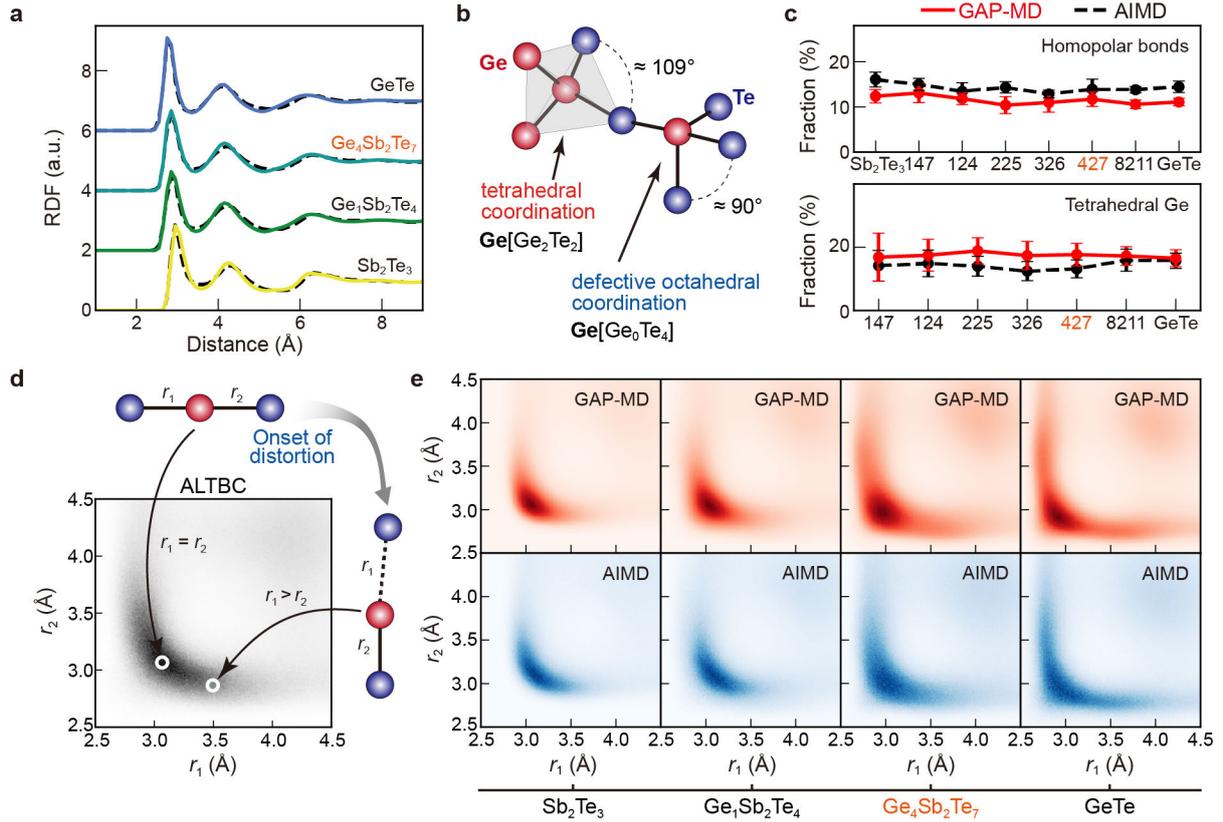

**Figure 2. Structures of disordered GST from ML-driven simulations.** (**a**) Radial distribution function (RDF) for four different compositions along the GeTe–$Sb_2Te_3$ compositional line, from GeTe (*cyan*) to $Sb_2Te_3$ (*yellow*). The $Ge_4Sb_2Te_7$ composition was not directly included in the training and is highlighted by red labels (*cf.* Supplementary Fig. 3). (**b**) Schematic sketch of octahedral and tetrahedral motifs in amorphous GST. (**c**) Fraction of homopolar bonds (Ge–Ge, Sb–Sb, Ge–Sb, and Te–Te) and tetrahedrally coordinated Ge atoms in simulated amorphous phases. Results from AIMD (GAP-MD) are shown in black (red), respectively. Error bars indicate standard deviations, and lines between points are guides to the eye. (**d**) Schematic sketch of the formation of a Peierls distortion, the degree of which can be quantified by the normalised angular-limited three-body correlation (ALTBC) function. (**e**) ALTBC functions for four amorphous GST phases as obtained from GAP-MD (*red*) and AIMD (*blue*), respectively.

The onset of a Peierls distortion is a more subtle structural effect in PCMs, which affects the size of the band gap in both crystalline and amorphous GST,[37] and is therefore directly relevant to applications in electronic memories. The degree of Peierls distortion can be assessed by comparing the short-to-long bond ratio in close-to linear environments (Fig. 2d), which we evaluate using the normalised angular limited three-body correlation (ALTBC) function. Figure 2e shows that our ML potential captures this effect very well: the ALTBC "fingerprints" for all four compositions studied are similar between GAP-MD (*red*) and the AIMD reference (*blue*).

We also carried out GAP-MD simulations with different stoichiometric compositions, "off" the GeTe–Sb$_2$Te$_3$ quasi-binary line, to account for the *p*-type semiconducting behaviour observed in experiments.[38] Specially, we considered two off-stoichiometric compositions, Ge$_2$Sb$_{1.8}$Te$_5$ (cation-deficient, *p*-type) and Ge$_{2.2}$Sb$_2$Te$_5$ (cation-rich, *n*-type), following previous AIMD work.[39] Structural indicators still yield close similarity between GAP-MD and AIMD (Supplementary Fig. 5): although the above compositions are slightly away from the training domain, their local bonding patterns were "learned" through sufficient sampling over structural and chemical space. All the above analyses demonstrate the compositionally transferable and defect-tolerant nature of the current ML model, enabling efficient MD simulations of amorphous GST with local compositional fluctuations—as they are expected to spontaneously occur upon repeated programming operations in memory devices.

**Towards device conditions.** Figure 3a shows three typical working conditions in which PCM thin films are produced by physical vapor deposition, such as magnetron sputtering; their thickness ranges from several to a few hundred nanometres. As-deposited thin films are usually amorphous and can be crystallised either by iso-thermal annealing or by Joule heating (induced by a programming pulse). In thin-film experiments, amorphous GST alloys crystallise into disordered rock-salt-like phases upon substrate heating at ≈ 150 ºC, accompanied by a density increase. Further annealing (250–400 ºC) leads to a gradual structural change towards the more ordered trigonal phase *via* continuous vacancy ordering.[38,40] Such annealing is carried out as the initial step for both electronic and photonic PCM devices for programming consistency. The subsequent operation in devices is to melt a fraction of the active volume near the heater by external electrical or laser pulses, and the following rapid cooling freezes the highly disordered structure to form the amorphous phase. In photonic devices that feature a more open structure, the active volume expands and shrinks frequently, whereas in electronic devices, the PCM is fully encapsulated within small memory cells, and the active volume remains nearly unchanged upon programming (Fig. 3a).

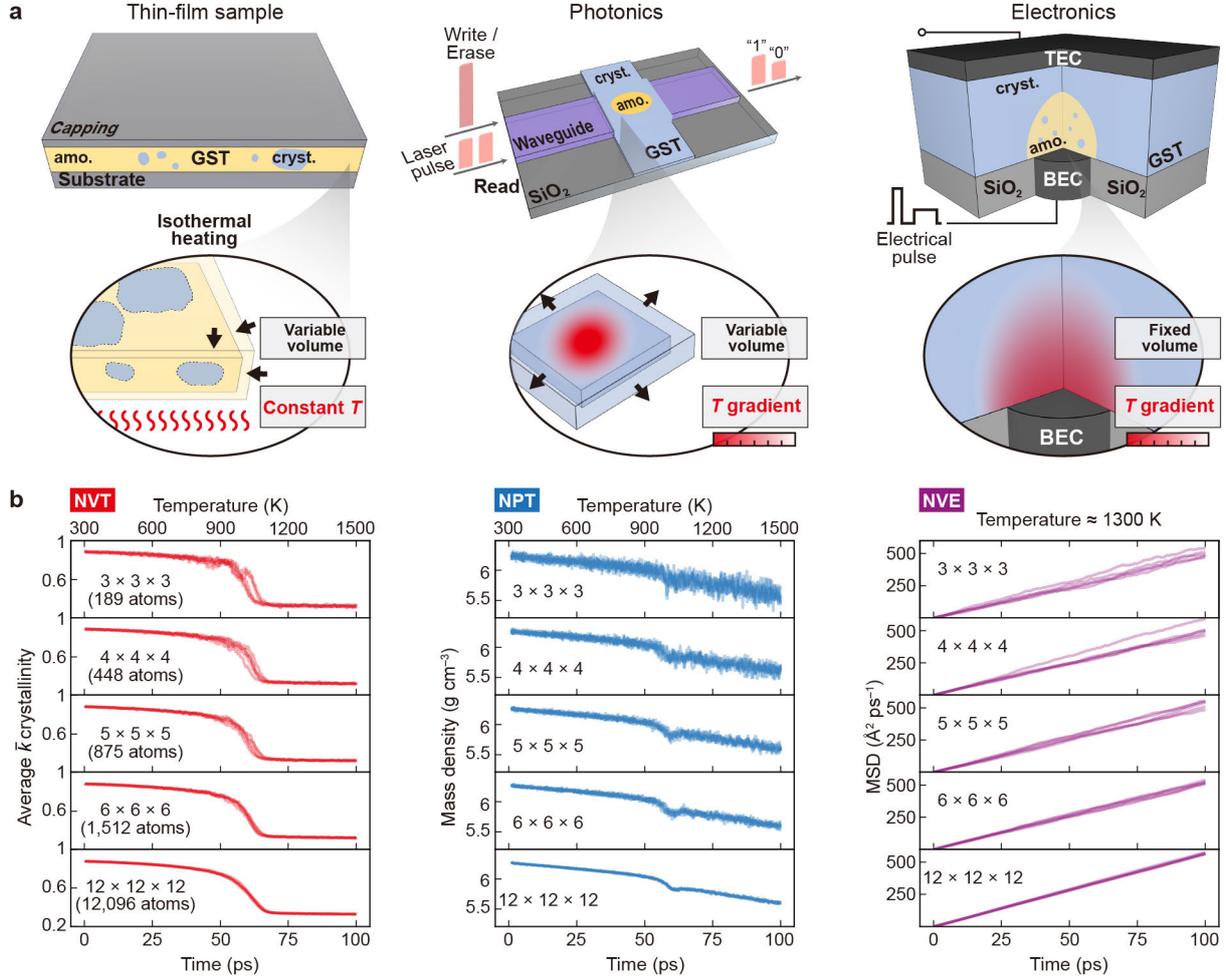

**Figure 3. Device conditions and modelling strategies.** (**a**) Schematic sketches showing, from left to right: thin-film laboratory samples, photonic devices,[10] and electronic devices.[9] Insets indicate different programming conditions in GST, the MD simulation of which requires different ensembles (NVT, NPT, and NVE). (**b**) Evolution of relevant quantities during GAP-MD simulations of $Ge_1Sb_2Te_4$ with different system sizes, expressed as $n \times n \times n$ expansions of the cubic rock-salt-type unit cell. From left to right, we show the per-cell averaged crystallinity measure, $\bar{k}$, during NVT heating, the mass density during NPT heating, and the mean square displacement (MSD) during NVE MD at 1,300 K (with initial configurations taken from the NVT runs). For each model size, five independent simulations were performed, and the results are plotted as semi-transparent lines.

To reflect these different situations, we tested different use cases for ML-driven MD: namely, simulations in the canonical (constant volume and temperature; NVT), the isobaric-isothermal (constant pressure and temperature; NPT), and the microcanonical (constant volume and energy; NVE) ensembles. Given the statistical distribution of vacancies in rock-salt-like GST and the highly disordered nature of liquid and amorphous GST, MD simulations of a few hundreds of atoms are naturally affected by strong finite-size effects. In parallel simulations of heating small-scale rock-salt-like $Ge_1Sb_2Te_4$ models (Fig. 3b), the SOAP-based crystallinity, $\bar{k}$,[19] falls at different onset temperatures for five independent runs with different initial atomic configurations (NVT ensemble), and large fluctuations in mass density are seen when the volume is allowed to change (NPT). In contrast to AIMD, GAP-MD enables efficient simulations for much larger model sizes, improving statistical sampling and thereby the description of the systems. In particular, the $\bar{k}$ and mass-density profiles nearly overlap for $12 \times 12 \times 12$ supercells (12,096 atoms each) over independent NVT and NPT runs. The same holds for NVE simulations, yielding more consistent liquid dynamics in larger supercells (Fig. 3b). The consistent structural and dynamical properties suggest marginal finite-size effects in larger supercells, as also supported by previous simulations of glassy $Ge_2Sb_2Te_5$.[41]

To simulate crystallisation (the SET operation) with AIMD, various intermediate densities or simply the crystalline density have been used for constant-volume simulations,[16-18] reflecting the highly confined geometric structure in electronic devices. Figure 4a shows a state-of-the-art AIMD crystallisation simulation of $Ge_1Sb_2Te_4$ over 600 ps at 600 K, with colour-coding for the per-atom similarity $\bar{k}$ (drawn with data from ref. [19]). The model contains 1,008 atoms (3.0 × 2.9 × 4.2 nm$^3$), with two fixed crystal-like layers representing the crystal–amorphous boundary encountered in devices (Fig. 3a). Our new GAP-MD simulation successfully reproduces the crystallisation process (Fig. 4b), and the resulting structure similarly contains a random distribution of vacancies and a number of anti-site defects. Note that this particular type of GAP-MD run can be completed within days on a high-performance computing system, whereas the corresponding AIMD simulation[19] required half a year of real time and shows much less favourable scaling behaviour.

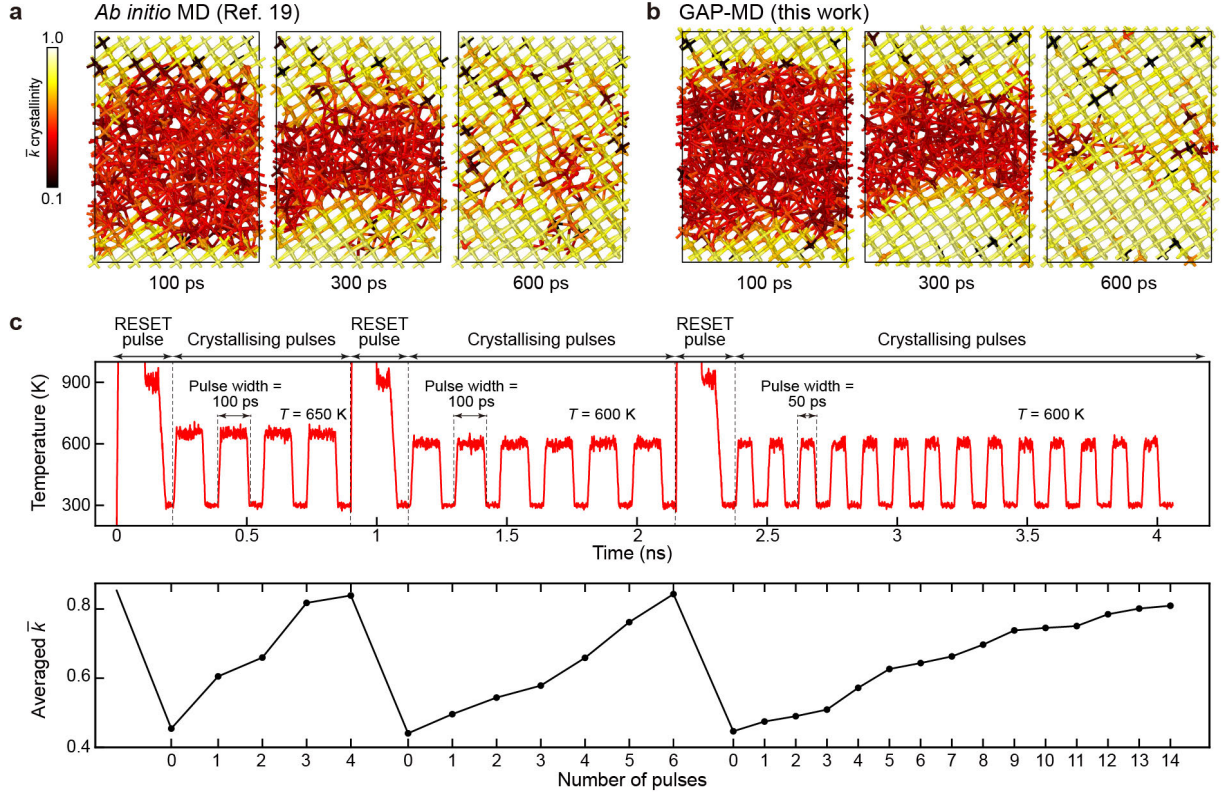

**Figure 4. Thermal cycles and cumulative SET operation.** (**a**) Snapshots from a 1,008-atom constant-temperature AIMD crystallisation trajectory for $Ge_1Sb_2Te_4$ in the ‹111› direction, taken from ref. [19] and visualised in the same style. Colour-coding indicates the SOAP-based $\bar{k}$ crystallinity,[19] illustrating the gradual transition from amorphous (*red*) to crystalline (*yellow*), and the occurrence of anti-site defects (*black*). (**b**) The same crystallisation simulation now carried out using GAP-MD. (**c**) GAP-driven simulations of cumulative SET processes. After each RESET (amorphising) pulse above the melting point of $Ge_1Sb_2Te_4$, a series of small SET (crystallising) pulses were applied, with the same simulation setup as shown in panel (b). In this proof-of-concept, three different kinds of pulses with different amplitudes and durations were considered: 650 K pulses of 100 ps duration, 600 K pulses of 100 ps duration, and 600 K pulses of 50 ps duration, respectively. The lower panel characterises the resulting cumulative SET processes by the averaged $\bar{k}$ crystallinity[19] as a function of the number of applied pulses. Each data point corresponds to a configuration after an individual pulse, and the lines are only guides to the eye; the starting point of the plot (corresponding to the time at 0 ns) refers to the fully recrystallised configuration from the simulation shown in panel (b).

**Single and cumulative switching.** In addition to SET and RESET, two further operations—cumulative SET and iterative RESET—are used for neuro-inspired computing.[4] In these cases, the programming pulse can be divided into a train of pulses with varied amplitude and duration,[42] and the change in electrical resistance[43] or optical transmission[44] as a function of the number of pulses can be used to mimic synaptic learning rules. Using the GAP ML model, we carried out cumulative SET simulations (Fig. 4c and Supplementary Movie 1). Differences in the initial configuration, pulse width, and maximum temperature all affect the progressive structural changes, measured here by a SOAP-based crystallinity indicator (as in Fig. 3 and in ref. [19]), shown in the lower panel of Fig. 4c. Note that the evolution of structural order determines electrical and optical properties of the intermediate states. These simulations represent the situation inside one grain of rock-salt-like GST (with the grain size varying from several to several tens of nanometres).[45,46] Taking into account the randomness in nucleation of GST on larger length scales, cumulative SET operations can emulate the integrate-and-fire dynamics for the construction of stochastic phase-change neurons.[47]

Simulating the iterative RESET process requires the use of the NVE ensemble in addition to NVT. We considered the widely used mushroom-type electronic device to model such an operation (Fig. 3a). To this end, we constructed a $32 \times 5 \times 20$ supercell expansion of a rock-salt-like $Ge_1Sb_2Te_4$ model, and applied energy pulses to an area described by a half-cylinder shape at the bottom of the *xz* plane (Fig. 5a). This setup represents the middle slab of mushroom cells, with a cross-section of about $20 \times 12$ nm$^2$. The top four atomic layers were fixed to prevent unwanted thermal transport to the top area through the periodic boundary conditions: in a real device, there would be a thermal barrier in contact with the GST material. We emphasise that in state-of-the-art mushroom-type devices,[46] the diameter of the heater can be scaled down to as small as ≈ 3 nm. Therefore, the expected effective heating area is directly mirrored by our simulation setup that is characterised in Fig. 5.

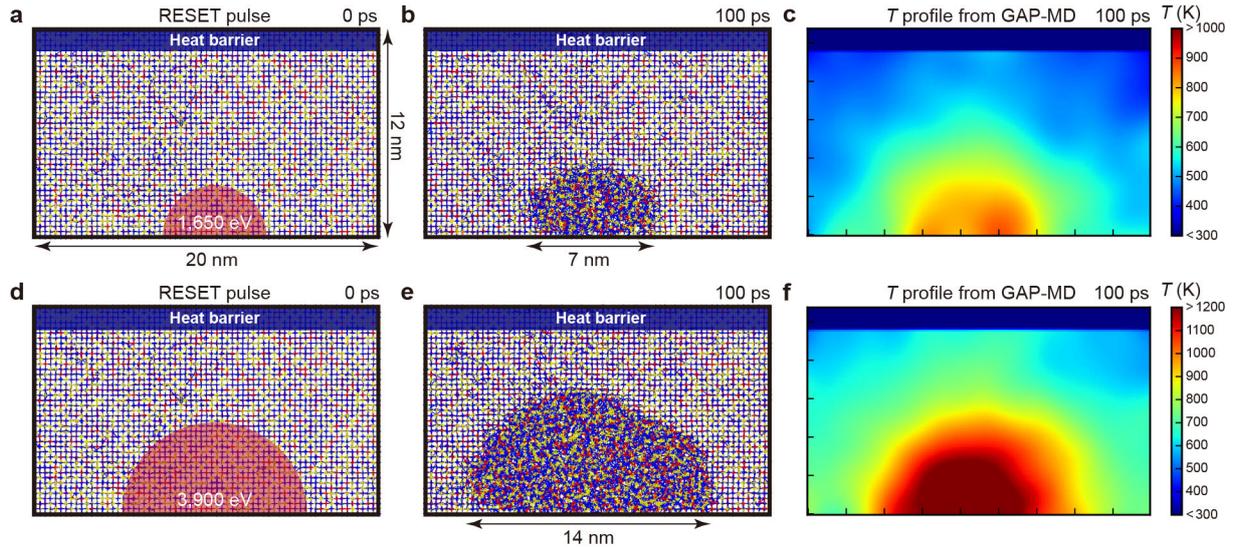

**Figure 5. Iterative RESET in a mushroom-type device setup.** (**a**) The initial rock-salt-like structure of $Ge_1Sb_2Te_4$, after being annealed at 300 K for 50 ps in the NVT ensemble. After that, an NVE simulation was carried out and an additional energy of 1,650 eV in total was imposed on the atoms as kinetic energy inside the focal area over the first 30 ps. (**b**) Snapshot of the NVE simulation at 100 ps, and (**c**) the corresponding two-dimensional temperature profile. The data were smoothed using a Gaussian filter with a broadness of 1.0 nm. (**d**–**f**) Same but for a heat pulse containing larger energy (3,900 eV in total). The snapshot and the corresponding temperature profile at 100 ps are shown.

After a short NVT calculation at 300 K, the ensemble was switched to NVE and additional energies of 55 eV ps$^{-1}$ were imposed on the atoms as kinetic energy inside the focal area over 30 ps, as described in the Methods section and indicated by the red area in Fig. 5a. The NVE simulation was continued for another 70 ps. In total, this 30 ps energy pulse added 1,650 eV (0.264 fJ) of thermal energy to the model, resulting in a disordered area with a base length of ≈ 7 nm (Fig. 5b). Then, we computed the local temperature profile based on the velocity of atoms collected on a two-dimensional grid. Figure 5c shows a clear radial temperature gradient from the heating centre towards the outer shell. In parallel, we performed another NVE calculation with an energy pulse of 3,900 eV (0.625 fJ) imposed on a larger focal area over 30 ps (Fig. 5d). The atomic structure and temperature profile after an NVE simulation over another 70 ps are shown in Fig. 5e and 5f. The stronger pulse generated a larger disordered area with a base length of ≈ 14 nm, giving rise to a more extended gradient profile with higher average temperature. This ML-driven MD simulation protocol closely resembles the iterative RESET operation in devices, providing real-time atomic-scale information beyond what has been revealed in experimental observations as well as finite-element method (FEM) simulations.[2] The evolution of the temperature gradient can be traced not only with sub-nanoscale spatial resolution, but also with an ultrafine time resolution at the femtosecond level (Supplementary Movie 2 and Supplementary Fig. 6).

**Unlocking full device-scale simulations.** Finally, we present a capability demonstration for a structural model of real device size, corresponding to a 3D Xpoint product as currently used in electronic memories.[5] Figure 6a shows the schematic layout of the crossbar array, which consists of a PCM memory layer (composition close to $Ge_1Sb_2Te_4$), an ovonic threshold switching (OTS) selector layer, and several buffer layers between PCM, OTS, and electrodes. The rest of the volume is filled with dielectric materials.[5] When programming, the electric current flows through the OTS layer after reaching a threshold voltage, and induces Joule heating to trigger the phase transition of the PCM. The size of the PCM region was reported as ≈ 20 × 20 × 40 $nm^3$ (ref. [5]). The initial state is taken to be the layered trigonal phase due to unavoidable thermal annealing during the backend-of-the-line (BEOL) stage for PCM integration.[2] We constructed a $Ge_1Sb_2Te_4$ supercell of the same size, containing more than half a million atoms in a layered configuration (30 septuple-layer blocks, with each block arranged in the stacking sequence Te–Sb–Te–Ge–Te–Sb–Te), and melted the model using a short (10 ps) but very-high-energy pulse with a high energy gradient. Such a short time window is experimentally accessible in principle: melting of GST can be achieved in 5–10 ps, as evidenced by femtosecond laser experiments.[48]

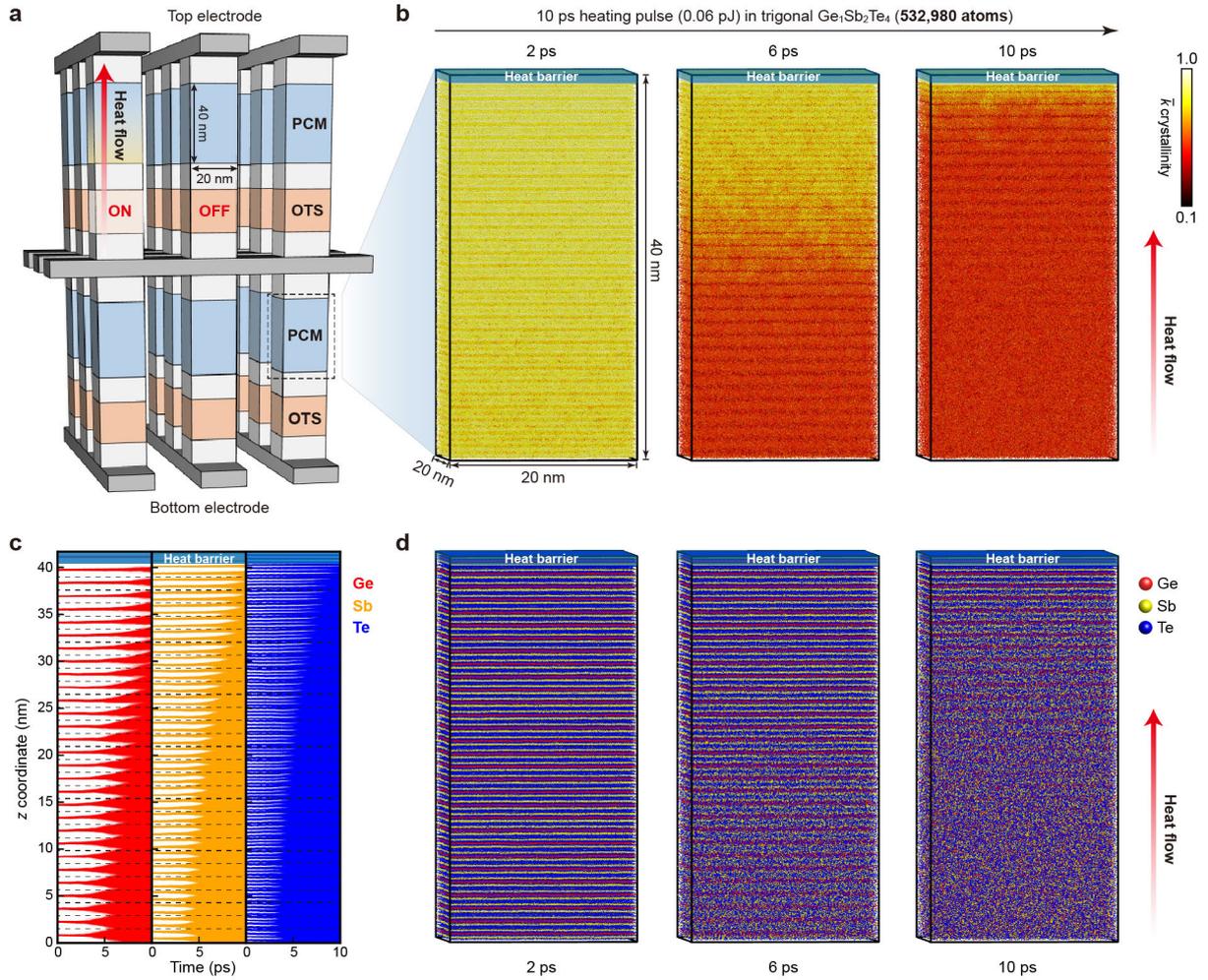

**Figure 6. Melting in a confined device setup.** (**a**) Schematic layout of commercial 3D Xpoint products, consisting of electrodes, PCM, OTS, and buffer layers in between, drawn based on information in ref. [5]. (**b**) Melting process of a $Ge_1Sb_2Te_4$ structural model that matches the dimensions of the PCM cell in 3D Xpoint memory (20 × 20 × 40 nm$^3$, containing 532,980 atoms). The melting process is triggered by a vertical heating pulse with additional energy (containing 60 fJ in total) from the bottom to the top after the cell is programmed. The colour coding is as in Fig. 4a–b, now indicating the gradual loss of crystalline order (*yellow*) upon formation of a disordered phase (*red*) in a much larger simulation cell. (**c**) Time-resolved evolution of the *z*-axis coordinates of the Ge, Sb, and Te atoms in the system. (**d**) As in panel (b), but now with colour-coding according to atomic species as in Fig. 1, emphasising the gradual melting of the layered GST structure.

Figure 6b illustrates the progressive disordering during melting. Layered structures remain early on (clearly seen in the plots 2 ps). Upon further heating, chemical disordering proceeded vertically from the bottom to the top, resulting in a gradient of ordering (Fig. 6b and Supple-mentary Movie 3). Having access to a fully atomistic simulation enables a more detailed analysis of the changes in the material structure:[49] plots of the individual $z$-axis coordinates provide a vertical profile of the evolution of the structure (Fig. 6c) and show that the melting of Ge layers starts slightly earlier than that of the Sb ones, and that the Te atoms near the van-der-Waals-like gaps are more mobile than those in the centre of the structural building blocks. At 6 ps, the bottom slab still shows some remaining chemical order, but it appears entirely disordered after 10 ps. The top region, especially the structural building blocks close to the heat barrier (near $z \approx 40$ nm), only partially melted after 10 ps.

The RESET energy is calculated as 60 fJ, which could be used as the theoretical limit of minimum power consumption for the initial operation per bit (the most power consuming step in devices). This is still an approximation because the entire amount of energy was directly assigned to the atoms with no energy dissipation to the surrounding layers. In practice, heat loss through the dielectric materials as well as electrodes increases the required RESET energy to the pico-Joule and even nano-Joule levels, calling for further improvements of thermal insulation.[2] Whilst the present capability demonstration focuses on the initial ground state, similar calculations and analyses can now be carried out for more complex structures, including realistic models of the rock-salt-like phase with grain boundaries or mixed crystalline / amorphous structures. Moreover, the simulated heating pulse can be adjusted to more realistic and complex programming conditions, considering various pulse amplitudes, durations and shapes, which shall help to guide the optimisation of programming schemes for better device performance.

## Discussion

Machine-learning models enable fully atomistic simulations of GST under realistic device geometry and programming conditions. The computational requirements of these types of ML-driven simulations increase linearly with the model system size,[41] and therefore they can be extended to even larger and more complex device geometries given the increasing availability of high-performance computing resources, in addition to longer timescales than in the present capability demonstrations. Beyond melting and crystal-growth simulations that are currently common in the field, we expect that our ML model will allow for a thorough sampling of nucleation and the atomic-scale observation of the formation of grain boundaries in large models of GST, either under isothermal conditions or with a temperature gradient. In combination with advanced sampling techniques,[18] ML-driven simulations should therefore allow to determine the nucleation barrier and critical nucleus size for GST. Moving from fundamental questions to those that are relevant to device engineering, in future work we envision the inclusion of interface effects on the surrounding electrode and dielectric layers. For example, it has been reported that heat loss can be largely reduced by confining the PCM cell with $Al_2O_3$ walls,[50] but their direct (atomic-scale) impact on thermal vibrations at the interface and on the phase-transition capacity of PCMs cannot be studied by FEM simulations alone. Atomistic ML models with extended reference databases are poised to provide this missing information, and in the future might help to predict the trend in minimum RESET energy as well as crystallisation time for different device geometries, thus enabling better architecture design. The computationally guided design not just of individual PCM phases, but of the entire devices based on them, has therefore come within reach.

**Methods**

**DFT computations.** DFT-based *ab initio* MD (AIMD) simulations used the "second-generation" Car–Parrinello scheme[51] as implemented in the Quickstep code of CP2K.[52] A combination of triple-zeta plus polarisation Gaussian-type basis set and plane waves with a cut-off energy of 300 Ry were used to expand the Kohn–Sham orbitals and the charge density, respectively. Scalar-relativistic Goedecker pseudopotentials were used.[53] Both the Perdew–Burke–Ernzerhof (PBE) functional[54] and a revised version for solids and surfaces (PBEsol)[55] were used in the present study. The Brillouin zone was sampled at Γ only, and the timestep was 2 fs.

Amorphous-phase structural models were generated following a standard melt-quench protocol.[56] Initial configurations were first randomised at 3,000 K (20 ps), then quickly cooled down to the corresponding melting points at a rate of $10^{14}$ K s$^{-1}$, and held there for 30 ps. To obtain amorphous structures, the models were then quenched to 300 K at $2.5 \times 10^{13}$ K s$^{-1}$, and held there for another 30 ps.

For the construction of the reference database, DFT reference values for atomic structures generated by AIMD and by iterative training (Fig. 1 and Supplementary Note 1) were computed with CASTEP[57] (version 16.11) using the PBEsol functional and ultrasoft pseudopotentials, yielding energies, forces, and stresses. A 600 eV cut-off for plane waves and an energy tolerance of $10^{-7}$ eV per cell were chosen for SCF convergence. A *k*-point grid of $2 \times 2 \times 2$ was used for large structural models (more than 150 atoms); other computations used an automatically generated *k*-point grid with a maximum spacing of 0.03 Å$^{-1}$.

To perform structural optimisation for crystalline GST phases using CASTEP, the plane-wave cut-off energy was set to 600 eV, and the energy tolerance of SCF convergence was $10^{-7}$ eV per cell. The criterion to break the ionic relaxation loop was a combination of an energy threshold of $10^{-6}$ eV per cell, a force tolerance of 0.01 eV Å$^{-1}$, and a stress tolerance of 0.05 GPa. A *k*-point grid with a spacing of 0.015 Å$^{-1}$ was used to sample reciprocal space.

**GAP fitting.** The ML interatomic potential was trained using the Gaussian approximation potential (GAP) framework,[35,58] using the QUIP and GAP codes (available for non-commercial research at https://github.com/libAtoms/QUIP). To avoid unphysical atomic clustering at high temperatures, we added a squared-exponential two-body (2b) term to the SOAP many-body descriptor,[36] with scaling factors of $\delta^{2b} = 1.0$ and $\delta^{SOAP} = 0.2$ used in the GAP fitting input. This combination of δ values led to a good description of individual dimer interactions (Supplementary Fig. 1). Details of SOAP and GAP hyperparameters used are given in Supplementary Note 2, and the numerical validation is described in Supplementary Note 3. The potential which is fitted to the iter2-3 database with PBEsol reference data is the primary version of the GST-GAP-22 model, and was used to generate the results shown in the main text.

We also provide a second version of the model that was "re-fitted" at the PBE level. Instead of repeating the entire iterative training process, we carried out single-point DFT computations using the PBE functional for the atomic structures included in the reference database. The obtained DFT energies, forces, and stresses were then used to fit the PBE version of the model. We verified that this version reproduces AIMD data at the PBE level with good accuracy, as shown in Supplementary Fig. 4. We note that there exists a difference in the computed lattice parameter (and thereby in the mass density), depending on which functional is used: PBEsol leads to a larger density than PBE. Nevertheless, in both cases, our GAP computations reproduce the DFT ones at the corresponding level (Supplementary Table 3).

**GAP-MD simulations.** The ability to carry out "device-size" simulations has long been relevant in atomistic modelling,[59] and ML-driven potentials are increasingly enabling very-large-scale simulations with millions of atoms and more.[60,61] In the field of PCMs, we mention previous simulations of GeTe nanowires with 16,540 atoms (ref. [62]) and of bulk amorphous $Ge_2Sb_2Te_5$ with 24,300 atoms in the simulation cell (ref. [41]). In the present work, we report GAP-driven MD simulations and link them directly to microscopic mechanisms in devices. These simulations were carried out using LAMMPS[63] with an interface to QUIP, using the canonical ensemble (NVT, constant number of particles, volume, and temperature), the isothermal–isobaric ensemble (NPT, constant number of particles, pressure, and temperature), and the microcanonical ensemble (NVE, constant number of particles, volume, and energy). The timestep was set to 2 fs (as in the AIMD runs). A Langevin thermostat was used to control temperature in the NVT simulations, and a Nosé–Hoover thermostat[64,65] controlling temperature and a barostat[66] controlling external pressure was used in the NPT simulations. Non-isothermal heating was described by NVE simulations. Additional energy was added to the kinetic energy of the atoms in the heating area, at a timestep of 2 fs.

**Structural analysis using SOAP.** The Smooth Overlap of Atomic Positions (SOAP) kernel measures the similarity of two atomic neighbour environments, quantified through their Gaussian-smoothed neighbour densities and a subsequent expansion in a set of local basis functions.[36] SOAP has been utilised for structural analysis for a variety of molecular and solid-state systems.[19,67-69] The SOAP kernel is defined for pairs of atoms, and the per-cell similarity can be constructed by averaging the descriptor vector entries over individual atomic environments in a given cell,[64] which is used to calculate and visualise the structural distances for the similarity map in Fig. 1b. The map was generated using t-distributed Stochastic Neighbor Embedding (t-SNE).[70]

To quantify the per-atom "crystallinity" (similarity to the corresponding crystalline phase) in melting and crystallisation simulations, each atom in a given MD snapshot was compared to all atomic environments of the same species (or the same group of elements, *viz.* Ge/Sb as cations) in a given reference model or models, and the SOAP similarities averaged to obtain what we call the "$\bar{k}$ crystallinity" (with $\bar{k} = 1$ indicating ideal crystal-like order), as in ref.[19]. As reference structural models, we considered three idealised rocksalt-like $Ge_1Sb_2Te_4$ models with randomised distributions of Ge atoms, Sb atoms, and vacancies on the cation-like sites (for the analyses shown in Fig. 3b and Fig. 4), as well as an idealised unit cell of trigonal $Ge_1Sb_2Te_4$ (for the device-scale simulation in Fig. 6).[45]

**Visualisation.** The sketch of the photonic device in Fig. 3a is based on ref.[71], the structural drawings in Fig. 1 were created using VESTA,[72] and those in Figures 4–6 were created using OVITO.[73]

## Data availability

Data supporting this work, including the parameter files required to use the potential, the fitting data, and the structural models shown in Figures 4–6, will be made openly available in a suitable repository upon journal publication. The XML identifier of the "iter2-3" potential model discussed in the main text is `GAP_2022_4_7_480_18_6_12_970`.

## Acknowledgements

W.Z. acknowledges useful discussions with Zhitang Song, Xudong Wang, and Riccardo Mazzarello. V.L.D. thanks Stephen R. Elliott for helpful comments. Y.Z. acknowledges support from a China Scholarship Council-University of Oxford scholarship. E.M. acknowledges the support of National Natural Science Foundation of China (52150710545). W.Z. acknowledges the support of 111 Project 2.0 (BP2018008) and the International Joint Laboratory for Micro/Nano Manufacturing and Measurement Technologies of Xi'an Jiaotong University. E.M. and W.Z. acknowledge XJTU for hosting their work at CAID. The authors would like to acknowledge the use of the University of Oxford Advanced Research Computing (ARC) facility in carrying out this work (http://dx.doi.org/10.5281/zenodo.22558), as well as computational resources by the HPC platform of Xi'an Jiaotong University and Hefei Advanced Computing Center. This work was performed using resources provided by the Cambridge Service for Data Driven Discovery (CSD3) operated by the University of Cambridge Research Computing Service (www.csd3.cam.ac.uk), provided by Dell EMC and Intel using Tier-2 funding from the Engineering and Physical Sciences Research Council (capital grant EP/T022159/1), and DiRAC funding from the Science and Technology Facilities Council (www.dirac.ac.uk). For the purpose of open access, the authors have applied a creative commons attribution (CC BY) licence to any author accepted manuscript version arising.

## Author contributions

W.Z. and V.L.D. designed and coordinated the research. Y.Z. developed the potential model with guidance from V.L.D., performed AIMD simulations with guidance from W.Z., and all other computations and analysis with guidance from all other authors. Y.Z., W.Z., and V.L.D. wrote the manuscript with input from E.M. All authors contributed to discussions.

## Competing interests

The authors declare no competing interests.

**Supplementary Note 1** (Components of the reference database)

The reference database contains several different parts which are listed below, and summarised in Supplementary Table 1.

**Free atoms and dimer data.** We included all three isolated atoms (Ge, Sb, and Te) in a 20 × 20 × 20 Å$^3$ cubic simulation cell. These cells provide energy references to fit. We also considered dimer configurations for all six different atomic pairs, *viz.* Ge–Ge, Ge–Sb, Ge–Te, Sb–Sb, Sb–Te, and Te–Te, again described using 20 × 20 × 20 Å$^3$ cubic simulation cells. For these dimer scans, we sampled interatomic distances between 1.0 and 7.0 Å, in which an interval of 0.1 Å was used from 1.0 to 2.0 Å and an interval of 0.2 Å from 2.0 to 7.0 Å.

**The initial "iter0" dataset.** The aim of this dataset is to facilitate the preliminarily exploration of different structural phases of GST, and to achieve a reasonable description of crystalline and disordered GST for the subsequent iterative rounds of training. The dataset consists of two parts, namely, crystalline GST structures and disordered configurations from AIMD trajectories.

In terms of crystalline phases, we considered the ground-state structures of the constituent elements, *viz.* Ge (diamond-type, $Fd\bar{3}m$), Sb (layered, $R\bar{3}m$), and Te (helices, $P3_121$). For seven GST compositions, we included both primitive cells and supercells of the stable (trigonal) phases, *viz.* $Sb_2Te_3$ ($R\bar{3}m$), $Ge_1Sb_4Te_7$ ($P\bar{3}m1$), $Ge_1Sb_2Te_4$ ($R\bar{3}m$), $Ge_2Sb_2Te_5$ ($P\bar{3}m1$), $Ge_3Sb_2Te_6$ ($R\bar{3}m$), and GeTe ($R3m$), and their metastable (rock-salt-like) phases in 2 × 2 × 2 or 3 × 3 × 3 supercell expansions, in which atomic vacancies are randomly distributed on the cation sites. In addition to the ideal crystalline forms of these crystals at varied (scaled) lattice parameters, we also added copies of the structural models with applied distortions to better sample the potential-energy surface (PES) of crystalline GST. Some cation-like atoms (Ge / Sb) were manually swapped with the anion-like ones (Te) to sample anti-site defects. We also included crystalline structures obtained from the Materials Project database,[S1] which include both the stable phases and certain metastable structures, with the goal to sample a more diverse range of configurations.

The disordered configurations in the "iter0" dataset also contain seven GST compositions, *viz.* $Sb_2Te_3$, $Ge_1Sb_4Te_7$, $Ge_1Sb_2Te_4$, $Ge_2Sb_2Te_5$, $Ge_3Sb_2Te_6$, $Ge_8Sb_2Te_{11}$, and GeTe, in cubic simulation cells containing fewer than 200 atoms, which were taken from AIMD melt-quench trajectories at different temperatures. We considered varied lattice parameters for these configurations, ranging from 17.7 Å to 19.2 Å with an interval of 0.3 Å, corresponding to commonly relevant atomic densities for frequently visited scenarios (that is, the highest density was chosen slightly larger than that of crystalline rock-salt-like GST, whereas the lowest is slightly smaller than the theoretical density of the amorphous phase at the PBE level[S2]).

**Standard iterations (iter1-1, iter1-2, and iter1-3).** The datasets obtained from standard iterations only include disordered configurations obtained from iterative GAP-driven melt-quench simulations. The objective of these three datasets is to extend the sampling of disordered configurations and thereby to enable the potential to accurately describe the structural properties of GST. We considered the same seven compositions as above ($Sb_2Te_3$, $Ge_1Sb_4Te_7$, $Ge_1Sb_2Te_4$, $Ge_2Sb_2Te_5$, $Ge_3Sb_2Te_6$, $Ge_8Sb_2Te_{11}$, and GeTe), and the various densities in the same range as in iter0. The models were generated following the same melt-quench protocol.

**Domain-specific iterations (iter2-1, iter2-2, and iter2-3).** Each iter2 dataset consists of three different kinds of configurations: corresponding to crystallisation, to melting, and to structures at even smaller (than the widely used ones) atomic densities. All three datasets include the seven compositions as in iter0 and iter1. To generate the intermediate configurations for the crystallisation process, we started from rocksalt-like structures of GST (with manually created anti-site defects as discussed in the iter0 dataset) and fixed two atomic layers as the crystalline matrix during the melt-quench process. The structural models were then annealed at 600 K without any geometrical constraints and gradually crystallised upon crystalline-amorphous interfaces. To obtain the melting configurations, the initial structures (*i.e.*, both rock-salt-like and trigonal GST) were heated up from 300 K to 1200 K without any geometric constrains. Intermediate configurations were taken from the heating process, in which different scales of atomic distortions were observed. Density variation was also considered, including commonly relevant values. The generation of configurations at even smaller atomic densities (than the commonly relevant ones) follows the same melt-quench protocol as discussed above.

**Supplementary Table 1.** Summary of the GST-GAP-22 reference database.

|  |  | Database size | | Sparse points |
|---|---|---|---|---|
|  |  | Cells | Atoms |  |
| Free atoms | Ge / Sb / Te | 3 | 3 | 1 |
| Dimer data | Ge–Ge / Ge–Sb / Ge–Te / Sb–Sb / Sb–Te / Te–Te | 210 | 420 | 90 |
| iter0 | Crystalline structures | 1261 | 69055 | 1500 |
|  | AIMD data | 210 | 41490 | 1200 |
| iter1-1 | Disordered (commonly relevant densities) | 112 | 21994 | 600 |
| iter1-2 |  | 112 | 21984 | 600 |
| iter1-3 |  | 112 | 21984 | 600 |
| iter2-1 | Crystallisation configurations | 84 | 21876 |  |
|  | Melting configurations | 84 | 21770 | 1200 |
|  | Disordered (more open structures) | 56 | 10992 |  |
| iter2-2 | Crystallisation configurations | 84 | 21876 |  |
|  | Melting configurations | 84 | 21914 | 1200 |
|  | Disordered (more open structures) | 56 | 10992 |  |
| iter2-3 | Crystallisation configurations | 84 | 21876 |  |
|  | Melting configurations | 84 | 21914 | 1200 |
|  | Disordered (more open structures) | 56 | 10992 |  |
| **Total** |  | **2692** | **341132** | **8191** |

## Supplementary Note 2 (GAP model fitting)

All potentials discussed in this work were fitted using the Gaussian Approximation Potential (GAP) framework,[S3] as implemented in the QUIP / GAP software. We refer the reader to a review article describing details of GAP and Gaussian process regression (ref. [S4]). In this note, we give details of the fitting process, and in Supplementary Note 3 below we discuss the numerical validation.

**SOAP and GAP hyperparameters.** We used a combination of square exponential two-body and SOAP many-body descriptors for the two-body (2b) and many-body terms in GAP. To combine the two, a weighting coefficient, $\delta$, is used to control the contributions from 2b and SOAP terms. To determine suitable values of $\delta$, we first fitted various potentials with different combinations of two $\delta$ values, based on the "iter0" datasets, and then performed dimer scans using these potentials (Supplementary Fig. 1). We found that models with either a 2b term only (that is, $\delta^{SOAP} = 0$) or a SOAP term only ($\delta^{2b} = 0$) cannot describe the dimer curves for the GST system well. Moreover, even though the combination of a small $\delta^{2b}$ (0.05) and a large $\delta^{SOAP}$ (0.2) led to a model which can well reproduce the region near the minimum of the dimer curves, it failed to accurately describe the interatomic interactions at small distances (< 1.0 Å). A rapid decay of the dimer curves when going to the small distances was found in this case, which might lead to atomic clustering once atoms are getting too close, especially at high temperatures. (This behavior is consistent with previous findings for a SOAP-only model for amorphous carbon, *cf.* ref. [S5]). Hence, we used a combination of a large $\delta^{2b}$ (1.0) and a small $\delta^{SOAP}$ (0.2) value in this work, and we show in Supplementary Fig. 1 that the resulting model can accurately describe the entire dimer scans in comparison to the DFT reference. (The values given here correspond to those used in the GAP fitting input.)

For the SOAP structural descriptor, we set the radial cut-off, $r_c$, to 5.5 Å, and the neighbour-density smoothness, $\sigma_{at}$, to 0.5 Å. For the neighbour-density expansion in local basis functions, we used a combination of $n_{max} = 12$ and $l_{max} = 4$, based on the previously described convergence of the accuracy of GAP models in terms of computational cost, as discussed in ref. [S4]. These hyperparameters were also used in the calculation of SOAP structural similarity measures for consistency (Fig. 1 in the main text). In the sparsification process (definition of "representative" points for the GAP model fit), we used different numbers of representative points for different parts of the dataset, according to the number of newly added configurations at each iteration. The objective of this scheme is to assign equivalent contributions from each iteration to the final potential, avoiding the overweighting of any specific subset. The numbers of representative points used are summarised in Supplementary Table 1.

## Supplementary Note 3 (Numerical validation)

To test the numerical accuracy of our GST-GAP-22 model, we performed both "internal" and "external" validation, based on configurations included in and excluded from the reference database, respectively. The results are summarised in Supplementary Table 2.

**Internal validation.** Five-fold cross validation was performed based on the reference database. We randomly split the database into 5 equivalent sub-sets ("folds"), and then predicted the cohesive energies and atomic forces of each fold (as the testing dataset), with the remaining folds as the training dataset. The reference database contains four main types of components, *viz.* crystalline phases, AIMD data, standard iterations, and domain-specific iterations. The computed cohesive energies (with respect to the isolated atoms) and force components using GAP are compared to the results obtained from DFT. Most datapoints scatter in a symmetric shape around the diagonal line (Supplementary Fig. 2a and 2b), which hints at a good numerical agreement between DFT and GAP models overall. We note that a limited number of crystalline configurations are not as well described by the model (corresponding to the datapoints away from the diagonal line) since they are either highly distorted crystals or rarely visited structures from the Materials Project database.[S1] Cumulative error distributions were calculated to assess the performance of our potential for different types of configurations. Almost 60% of the crystalline configurations have an absolute cohesive energy error of less than 0.01 eV atom$^{-1}$, and for standard and domain-specific iterations this value increases to about 80% and about 90%, respectively. Most energy errors for AIMD-based structures range from 0.01 eV atom$^{-1}$ to 0.1 eV atom$^{-1}$, slightly larger than for the other three. The cumulative force error plots (indicating the overall distribution) are rather similar for the four parts of the database, consistent with similar RMSEs (*cf.* Table 1 in the main text).

**External validation.** We also carry out an out-of-sample validation, in which the testing configurations were taken from AIMD trajectories that were not included in the training datasets. These testing databases consist of two parts. The first one contains 80 disordered GST structures (including $Sb_2Te_3$, $Ge_1Sb_4Te_7$, $Ge_1Sb_2Te_4$, $Ge_2Sb_2Te_5$, $Ge_3Sb_2Te_6$, $Ge_4Sb_2Te_7$, $Ge_8Sb_2Te_{11}$, and GeTe, with each simulation cell containing more than 300 atoms) obtained from melt-quench AIMD simulations. These results of AIMD simulations were also used as a reference in the structural analysis (Fig. 2 in the main text). The second part contains intermediate configurations (1,008 atoms per cell) of the recrystallisation process of $GeSb_2Te_4$ as described using AIMD simulations in ref. [S6], with data taken from that previous work.[S6] We compared the energies and forces using our GAP potential to the AIMD reference (Supplementary Fig. 2c). All energy errors for these out-of-sample structures are lower than 0.5 eV atom$^{-1}$. Most forces errors are about 0.1 eV Å$^{-1}$, with some values being as small as $10^{-4}$ eV Å$^{-1}$.

**Supplementary Table 2.** Root mean square error (RMSE) results for predicted energies and force components, obtained through both internal (cross-validation) and external (out-of-sample) testing.

|  |  | RMSE energies (eV atom$^{-1}$) | RMSE forces (eV Å$^{-1}$) |
|---|---|---|---|
| Cross-validation | Distorted crystals | 0.024 | 0.20 |
|  | Disordered AIMD | 0.027 | 0.18 |
|  | Standard iterations | 0.010 | 0.18 |
|  | Domain-specific iterations | 0.007 | 0.18 |
| Out-of-sample (AIMD) | Disordered structures (> 300 atoms; this work) | 0.023 | 0.16 |
|  | Crystallisation trajectories (1,008 atoms; ref. [6]) | 0.025 | 0.15 |

**Supplementary Table 3.** Theoretical lattice parameters of crystalline GST, $Sb_2Te_3$, and GeTe in their ground state obtained by both GAP-MD and DFT computations at 0 K. The difference in lattice parameters stems from the use of different functionals in the DFT computations.

|  |  | PBEsol | | | PBE | | |
|---|---|---|---|---|---|---|---|
|  |  | GAP | DFT | Error (%) | GAP | DFT | Error (%) |
| $Sb_2Te_3$ | $a$ (Å) | 4.252 | 4.275 | 0.5 | 4.345 | 4.339 | 0.1 |
|  | $c$ (Å) | 30.114 | 29.786 | 1.1 | 31.006 | 31.231 | 0.7 |
| $Ge_1Sb_4Te_7$ | $a$ (Å) | 4.253 | 4.257 | 0.1 | 4.331 | 4.323 | 0.2 |
|  | $c$ (Å) | 23.322 | 23.278 | 0.2 | 24.053 | 24.326 | 1.1 |
| $Ge_1Sb_2Te_4$ | $a$ (Å) | 4.254 | 4.244 | 0.2 | 4.321 | 4.312 | 0.2 |
|  | $c$ (Å) | 39.867 | 40.033 | 0.4 | 41.054 | 41.782 | 1.7 |
| $Ge_2Sb_2Te_5$ | $a$ (Å) | 4.245 | 4.226 | 0.5 | 4.310 | 4.298 | 0.3 |
|  | $c$ (Å) | 16.701 | 16.780 | 0.5 | 17.107 | 17.396 | 1.7 |
| GeTe | $a$ (Å) | 4.173 | 4.161 | 0.3 | 4.242 | 4.227 | 0.4 |
|  | $c$ (Å) | 10.355 | 10.448 | 0.9 | 10.698 | 10.895 | 1.8 |

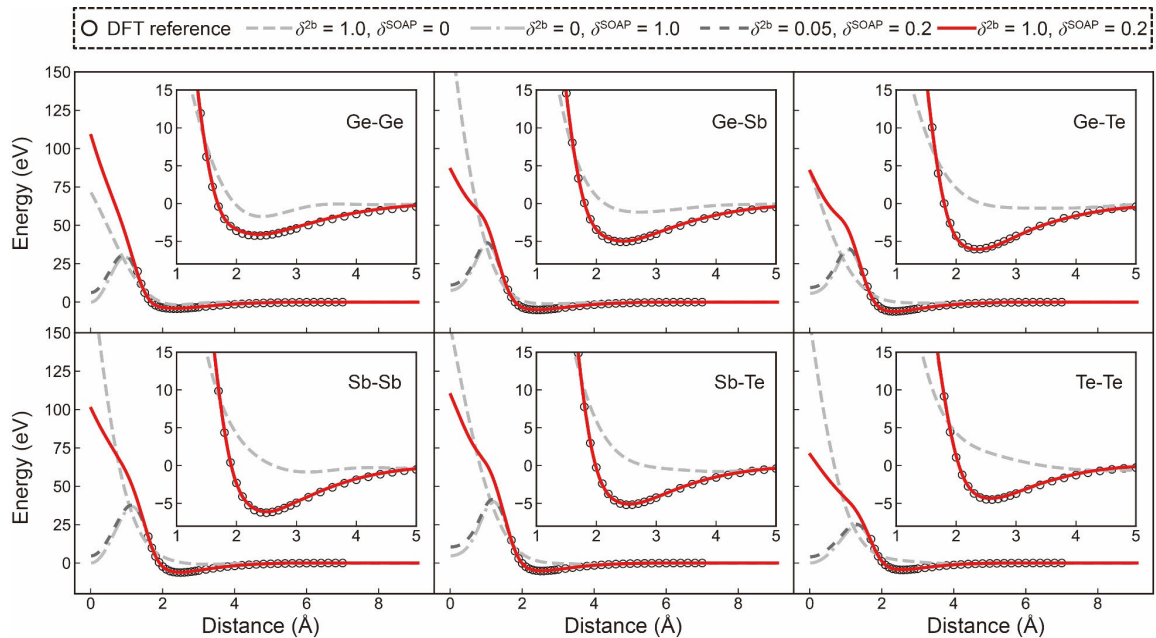

**Supplementary Figure 1.** Results of isolated-dimer scans relevant to the Ge–Sb–Te system, including Ge–Ge, Ge–Sb, Ge–Te, Sb–Sb, Sb–Te, and Te–Te. Energies relative to the free atoms were computed from various GAP models with different $\delta$ values and compared to DFT data.

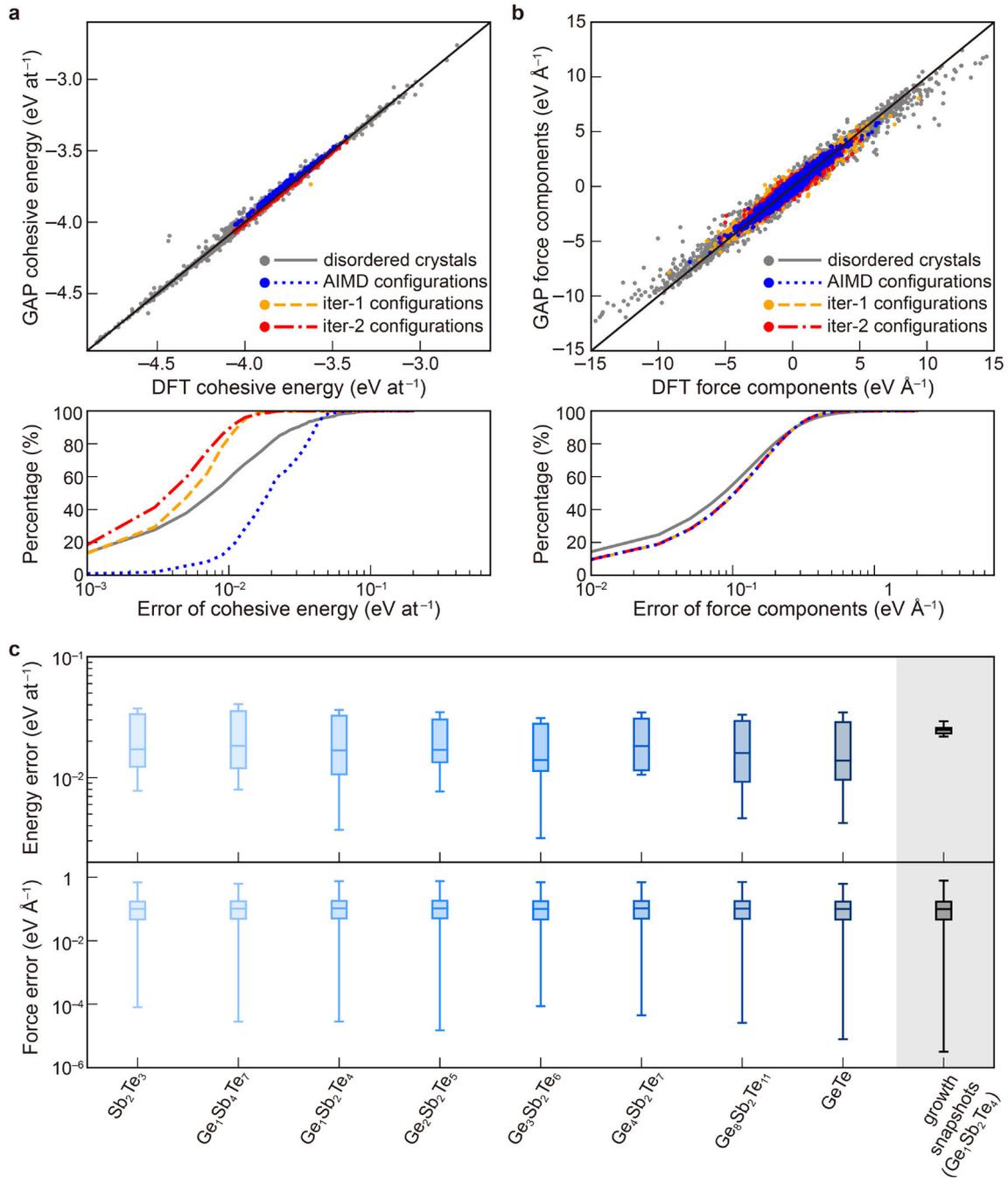

**Supplementary Figure 2.** (**a**) Scatterplot of cohesive energies, relative to the constituent free atoms, for configurations of different types. The ground-truth DFT value is shown on the horizontal axis, and the prediction of the ML model is shown on the vertical axis. A cumulative error plot, indicating the percentage of data points that have, at maximum, a given error, is given below. (**b**) Same for errors of the Cartesian force components. (**c**) Box plots of the errors of cohesive energies and forces, for AIMD configurations of various disordered GST phases (*left*) and AIMD growth simulations of Ge$_1$Sb$_2$Te$_4$ (*right*). Structural data for the former were generated for the present work, whereas structural data for the latter were taken from ref. [S6]. In the box plots, the boxes indicate the range from 25% to 75% with a horizontal line indicating the median, and the whiskers represent the full range of the data.

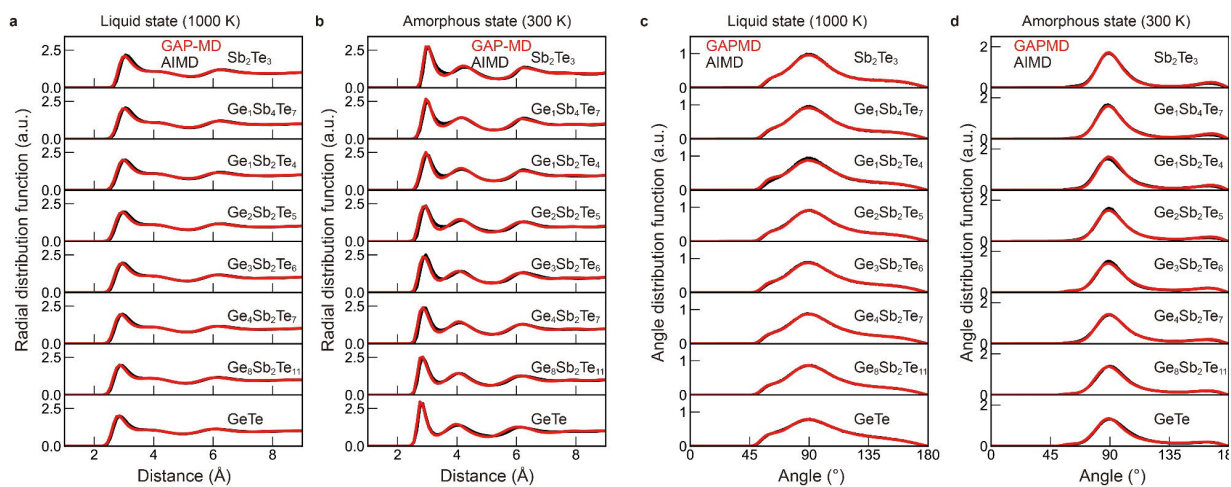

**Supplementary Figure 3.** Calculated radial distribution function (RDF) and angular distribution function (ADF) plots, obtained using GAP-MD (PBEsol) and AIMD (PBEsol). Data are shown for 8 compositions on the quasi-binary line, *viz.* $Sb_2Te_3$, $Ge_1Sb_4Te_7$, $Ge_1Sb_2Te_4$, $Ge_2Sb_2Te_5$, $Ge_3Sb_2Te_6$, $Ge_8Sb_2Te_{11}$, and GeTe which were included in the training database, and $Ge_4Sb_2Te_7$ which was not included.

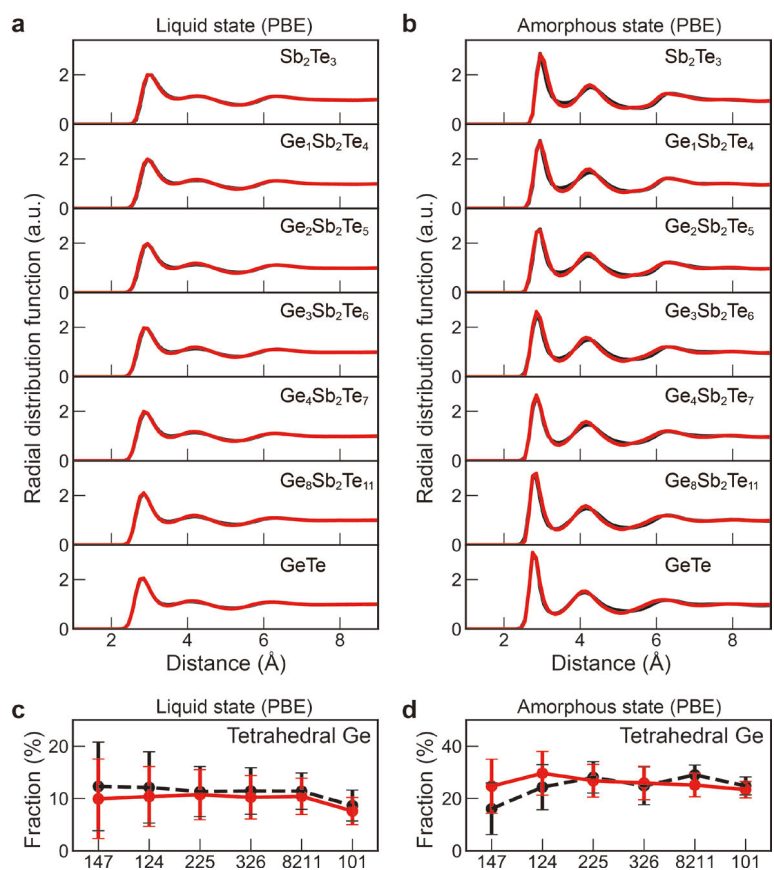

**Supplementary Figure 4.** Structural validation for a version of the GAP model fitted using PBE reference data. (**a**–**b**) Computed radial distribution function (RDF) plots for liquid and amorphous GST structures obtained from GAP-MD in comparison to AIMD, both based on the PBE functional. (**c**–**d**) The fraction of tetrahedrally coordinated Ge atoms in liquid and amorphous GST, calculated from GAP-MD (PBE) in comparison to AIMD (PBE) data.

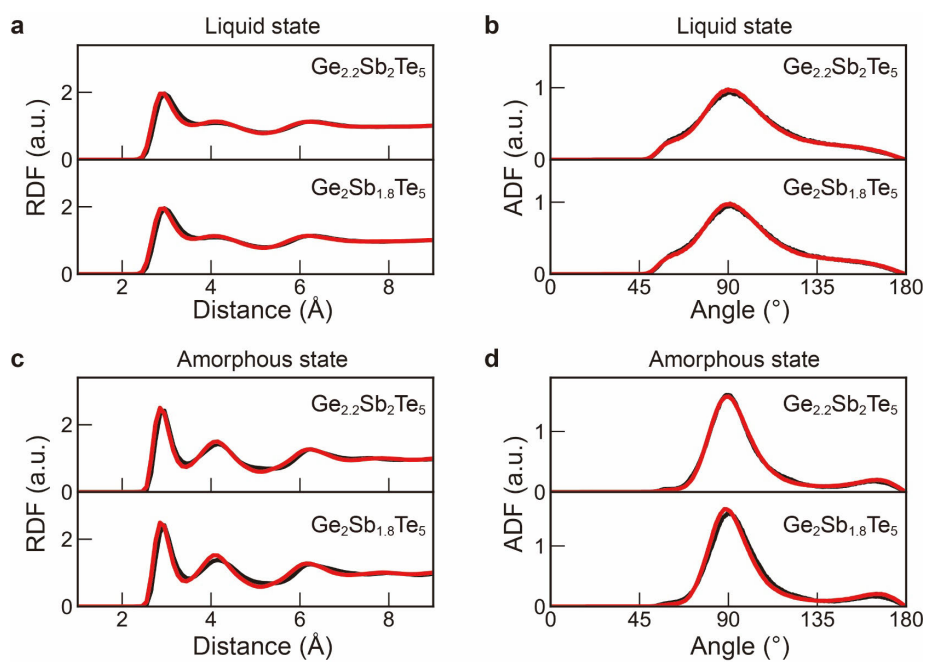

**Supplementary Figure 5.** Structural validation for "off-stoichiometric" compositions (compounds not included on the quasi-binary line) as described by GAP-MD (PBEsol) compared to AIMD (PBEsol) data. (**a**–**b**) Radial distribution function (RDF) and angular distribution function (ADF) of liquid $Ge_{2.2}Sb_2Te_5$ and $Ge_2Sb_{1.8}Te_5$, respectively. (**c**–**d**) Same for the corresponding amorphous phases.

S10

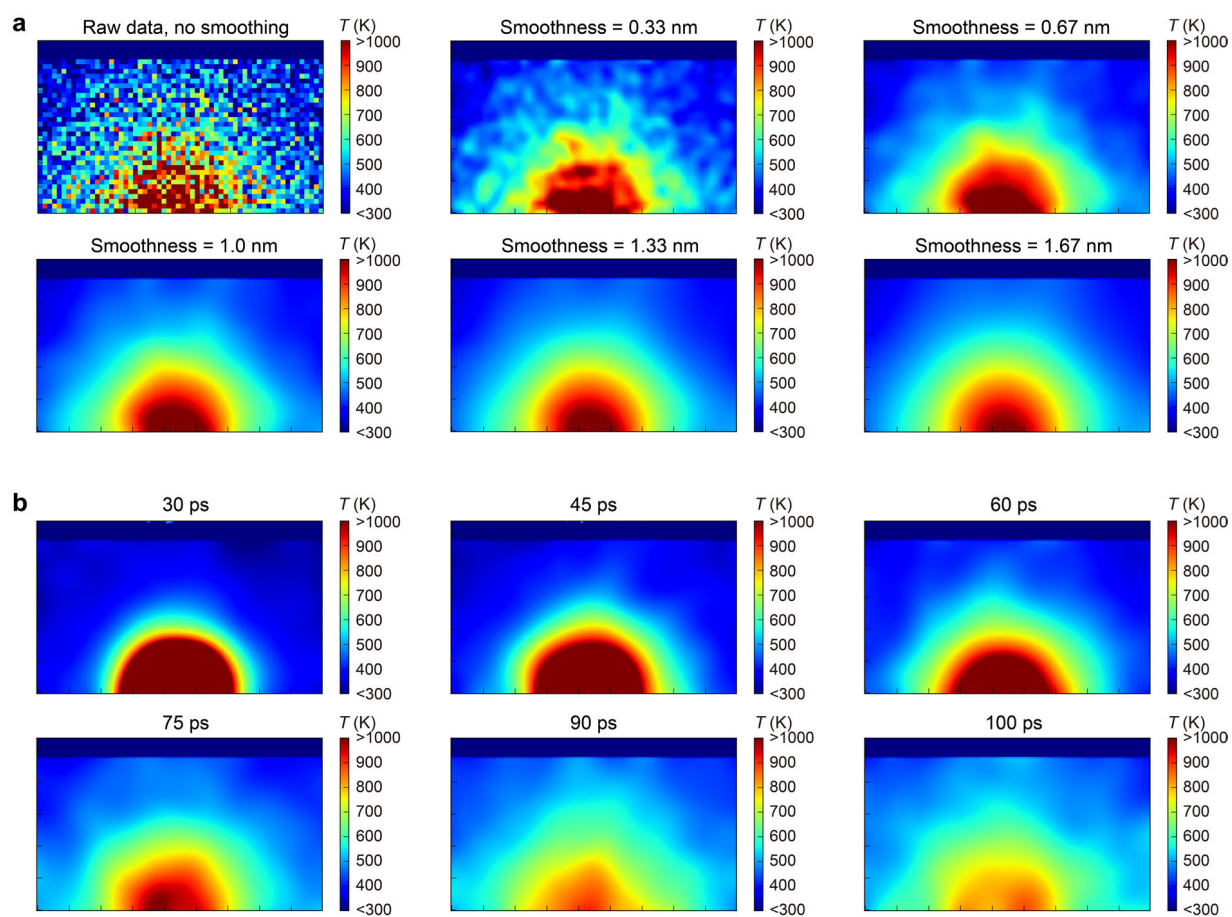

**Supplementary Figure 6.** The calculated temperature profile of the mushroom model (Fig. 5a in the main text) during the non-isothermal heating process. (**a**) The cross-section (perpendicular to the plane shown) of the model was first divided into a 60 × 36 grid (21,600 pixels in total). The per-atom temperature was calculated as an average over all atoms in each grid pixel, leading to the raw data shown in the top left panel. Gaussian filters with different smoothness values can be used to smear out the statistical fluctuation of calculated temperatures. After smoothing, the resolution was further improved by linear interpolation using a finer grid. (**b**) Snapshots of the evolution of the temperature gradient during the non-isothermal heating process. After a 30-ps heating pulse, another 70-ps NVE simulation was performed. A Gaussian filter with a smoothness of 1.0 nm was applied to the data. The full process is shown in Supplementary Movie 2.

## Supplementary References